# Time Reversal Enabled Fiber-Optic Time Synchronization

Yufeng Chen, Hongfei Dai, Wenlin Li, Fangmin Wang, Bo Wang, and Lijun Wang

*Abstract*—Over the past few decades, fiber-optic time synchronization (FOTS) has provided fundamental support for the efficient operation of modern society. Looking toward the future beyond fifth-generation/sixth-generation (B5G/6G) scenarios and very large radio telescope arrays, developing high-precision, low-complexity and scalable FOTS technology is crucial for building a large-scale time synchronization network. However, the traditional two-way FOTS method needs a data layer to exchange time delay information. This increases the complexity of system and makes it impossible to realize multiple-access time synchronization. In this paper, a time reversal enabled FOTS method is proposed. It measures the clock difference between two locations without involving a data layer, which can reduce the complexity of the system. Moreover, it can also achieve multiple-access time synchronization along the fiber link. Tests over a 230 km fiber link have been carried out to demonstrate the high performance of the proposed method.

*Index Terms*—Fiber-optic time synchronization, time reversal, scalability, multiple access, fifth-generation/sixth-generation, time deviation.

## I. INTRODUCTION

2022 is declared as the International Year of Glass by the United Nations, which shows the passion of celebrating the glass's (fiber's) essential role in modern society and making the best use of fiber in different areas [1]. As an important application of fiber, fiber-optic time synchronization (FOTS) has played a fundamental role in a wide range from civil, military, to science applications [2]-[16]. Compared with satellite-based time synchronization, FOTS has a unique advantage in accuracy [2], owing to the superior stability and transmission reciprocity of the fiber. Moreover, its good immunity to electromagnetic interference makes FOTS a suitable timing solution in scenarios where electromagnetic silence is needed or electromagnetic interference is high. For scientific infrastructures such as particle accelerators, the distributed sensors and actuators need to be coordinated precisely in timing for controlling and monitoring the beam [3], [4]. In radio astronomy, FOTS can provide time reference for telescope arrays to precisely record the arrival time of radio signals at each receptor [5], [6]. More specifically, FOTS has already been applied in astronomic facilities such as the Atacama Millimetre Array (ALMA) [7], the e-MERLIN [8] and the Square Kilometre Array (SKA) [9], [10] and will be implemented in future radio telescope arrays, such as next-generation Very Large Array (ngVLA) [11]. In addition, FOTS has been widely used in other fields such as deep-space communication and tracking [12], geodesy [14], [15], navigation as well as telecommunication [16].

Besides, FOTS is playing an increasingly important role in the Internet of Things (IoT), smart cities and grids, distributed ledger technologies and multidimensional sensing services [17], [18], [19], [20]. In future fifth-generation/sixth-generation (B5G/6G) scenarios, building a space-air-ground integrated network (SAGIN) can provide full-area three-dimensional coverage and broadband access capabilities anytime, anywhere for anyone [21], [22]. As can be expected, such an integrated network puts forward higher demands on the scalability, high precision and low complexity of the time synchronization system. Scalability can make it feasible to build a large-scale time synchronization network. Meanwhile, low complexity tends to bring the advantages of low cost, low power consumption, and low technical threshold for installation and maintenance. Therefore, it is foreseeable that developing high-precision, low-complexity and scalable FOTS technology will be vital to the B5G/6G integrated network.

At present, the commonly used FOTS scheme is two-way time transfer [23], [24], [25], [26], [27]. Both the server site (with time reference) and the user site (to be synchronized to the server site) have their own clocks. To compensate the clock offset of the two sites, the measured time difference between the received time signal and the local time signal should be exchanged in real time. As a result, the data layer is needed, and encoder/decoder are also required at both sites. It increases the complexity of the system and theoretically limits the accuracy of time synchronization. Furthermore, since the timing data in the traditional two-way FOTS scheme are specific to the two ends, it is impossible to realize multiple access functions [28]. This greatly limits the scalability of the system.

In this paper, we propose a time reversal enabled FOTS method, which has the characteristics of high precision, scalability and low complexity. The concept of time reversal has been applied in the fields of biomedical ultrasound imaging and therapy [29], underwater acoustics [30], and nondestructive testing [31]. But to the best of our knowledge, it has never been

This work was supported in part by the National Natural Science Foundation of China under Grant 61971259, in part by National Key R&D Program of China under Grant 2021YFA1402102, and in part by Tsinghua University Initiative Scientific Research Program. *(Corresponding author: Bo Wang).* These authors contributed equally: Yufeng Chen, Hongfei Dai.

The authors are with Department of Precision Instrument, Tsinghua University, Beijing 100084, China and also with the Key Laboratory of Photonic Control Technology (Tsinghua University), Ministry of Education, Beijing 100084, China (e-mail: bo.wang@tsinghua.edu.cn).

implemented in fiber-based time synchronization before. Time reversal provides a new idea for time synchronization: the clock offset can be compensated without calculating the fiber link delay. The advantages derived from time reversal are as follows. First, enabled by the time reversal operation, the fiber propagation delay fluctuation in two directions can be directly eliminated. Hence, the data layer is no longer needed during transmission, which can effectively reduce the complexity of the system. Second, the synchronization process of this method is very simple, requiring only two steps: synchronization request and synchronization response. This process is a direct and efficient way to achieve time synchronization. It can provide the possibility of synchronization for multiple remote user sites, with the help of time-division multiplexing technology and fast switchable optical switches. Third, since the physical time signal instead of the encoded timing data is transmitted over the fiber link, it is feasible to achieve multiple-access time synchronization.

To verify the performance of the proposed time reversal enabled FOTS method, experiments are carried out on a 230 km fiber link. The time synchronization result characterized by time deviation (TDEV) of ~2 ps is achieved at an averaging time of 1000 s. To demonstrate its multiple-access capability, a point 50 km away from the server site is selected to recover the time signal. TDEV of ~3 ps at an averaging time of 1000 s is obtained.

## II. PRINCIPLE

Time reversal is an operation of reversing the direction of time based on the transform: $t \to -t$ or $t \to C - t$, where $t$ represents the time and $C$ is a constant [32], [26]. As shown in Fig. 1(a), time reversal of a time signal can be understood as adding a negative time delay and making time run backward. However, since a negative time delay ($-t$) cannot be realized, a practical method for time reversal is to generate a positive delay in the form of $C - t$ ($C$ is a constant delay). The simplified timing model is shown in Fig. 1(b). The upper and lower black axes indicate relative time relationships within one period. The time signals at the server site and user site are labeled $t_{server}$ and $t_{user}$, respectively. There exists a clock offset $T_{offset} = t_{server} - t_{user}$. The key step is to measure this clock offset and maintain it as a known constant value.

To obtain the clock offset, the time signal at the user site is sent to the server site through a fiber link. This process is viewed as synchronization request (*Sync_Req*). At the server site, a time interval counter (TIC) is used to measure the time difference $T_1$ between the local time signal $t_{server}$ and the received time signal $t_{rxs}$:

$$T_1 = t_{rxs} - t_{server} = \tau_{u-s} - T_{offset}, \quad (1)$$

where $\tau_{u-s}$ is the one-way propagation delay of the fiber link from the user site to the server site. To implement the time reversal operation, the time signal $t_{server}$ is delayed by $C - T_1$ ($C > T_1$), and moved to the position of $t_{s-reversal}$, which can be expressed as

$$t_{s-reversal} = t_{server} + C + T_{offset} - \tau_{u-s}. \quad (2)$$

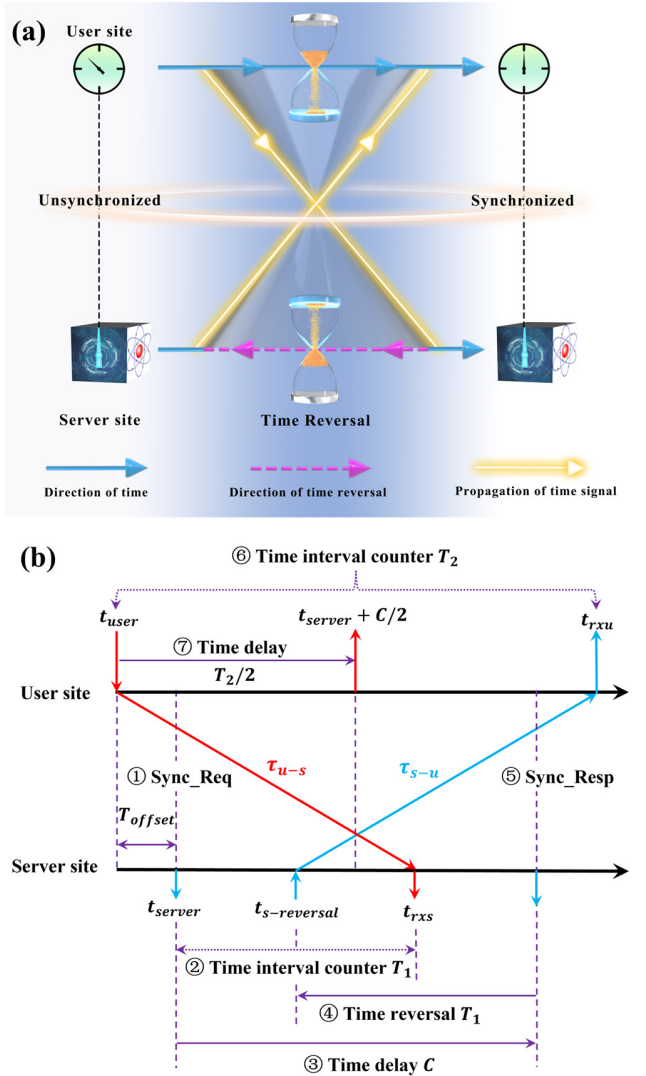

Fig. 1. Schematic diagram of time reversal enabled FOTS. (a) Conceptual diagram of time reversal. To be time synchronized to the server time, the user time signal is sent from user site to server site. At the server site, the server time signal is reversed by the time difference between the received user time signal and the local server time signal. Subsequently, the reversed server time signal is sent to the user site. Finally, time synchronization can be established at the user site by measuring the time difference between the received reversed server time signal and the local user time signal. (b) Simplified timing model of the time reversal enabled FOTS method. It includes synchronization request (*Sync_Req*) process from user site to server site and synchronization response (*Sync_Resp*) process from server site to user site. The upper and lower black axes indicate relative time relationships within one period.

Then, the synchronization response (*Sycn_Resp*) process is carried out. At the server site, the reversed time signal $t_{s-reversal}$ is sent back to the user site through the same fiber link. The received time signal $t_{rxu}$ at the user site can be expressed as

$$t_{rxu} = t_{server} + C + T_{offset} - \tau_{u-s} + \tau_{s-u}, \quad (3)$$

where $\tau_{s-u}$ is the one-way propagation delay of the fiber link from server site to user site. Due to the reciprocity of the fiber link, the value of $\tau_{s-u} - \tau_{u-s}$ can be regarded as 0, or at least a constant that can be calibrated considering the Sagnac effect and propagation delay asymmetry in practical applications. The detailed calibration process can be found in the Appendix.

At the user site, the time difference $T_2$ between the local time signal $t_{user}$ and the received time signal $t_{rxu}$ can be measured and expressed as:

$$T_2 = t_{rxu} - t_{user} = C + 2T_{offset}. \qquad (4)$$

Delaying the time signal $t_{user}$ by $T_2/2$, the time signal $t_{server} + C/2$ can be obtained at the user site. Since $C$ is a known constant value, time synchronization is realized.

## III. EXPERIMENTAL SETUP AND RESULTS

*A. Time reversal enabled time synchronization on a 230 km fiber link*

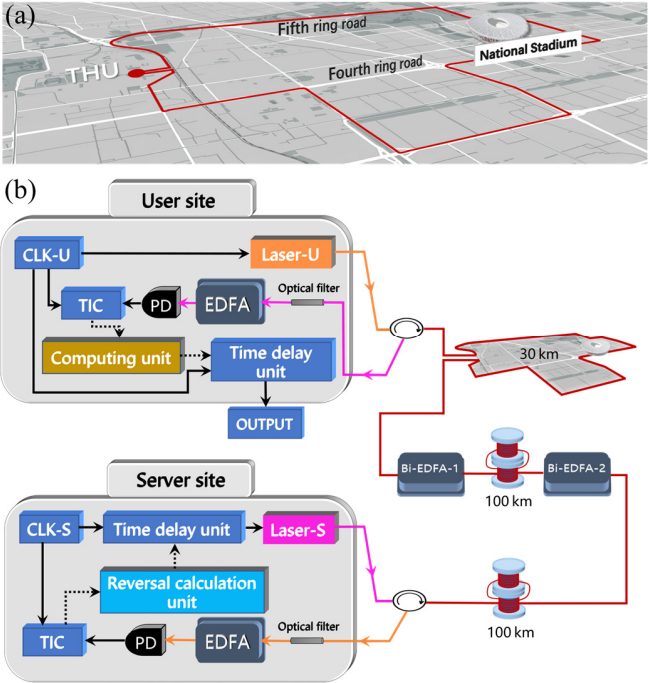

Fig. 2. The experimental diagram of the time reversal enabled FOTS over a 230 km fiber link, which is made up of 30 km urban fiber link in Beijing and 200 km fiber spools in the lab. (a) Schematic diagram of the 30 km urban fiber link. It starts from Tsinghua University (THU), surrounds the Olympic center and the Bird's nest (National Stadium), and returns to THU at last. (b) Schematic diagram of the experimental setup. The server site and user site are located in the same lab. At both sites, the orange lines represent the optical signal with a wavelength of 1546.92 nm (laser-U), and the pink lines represent the optical signal with a wavelength of 1546.12 nm (laser-S). CLK-U, clock signal at user site; TIC, time interval counter; PD, photodetector; EDFA, erbium-doped fiber amplifier; Bi-EDFA, bidirectional erbium-doped fiber amplifier; CLK-S, clock signal at server site.

The experimental diagram of the time reversal enabled FOTS is shown in Fig. 2. The server site and user site are connected by a 230 km fiber link, which is made up of 30 km urban fiber link in Beijing and 200 km fiber spools in the lab. As shown in Fig. 2(a), the urban fiber link starts from Tsinghua University (THU), surrounds the Olympic center and Bird's nest (National Stadium), and returns to THU at last. For convenience, server and user sites are both located in the same lab at THU. The detailed setup is shown in Fig. 2(b). At the user site, one pulse per 10 milliseconds signal CLK-U generated by the clock generator (model CG635, Standard Research Systems Corporation) is used to modulate a distributed feedback (DFB) laser (laser-U), whose wavelength is 1546.92 nm. The optical signal propagates through the 230 km fiber link, optical filter and erbium-doped fiber amplifier (EDFA), and is detected by a photodetector (PD) at the server site. This process corresponds to *Sync_Req* in Fig. 1(b). The time difference measured by a TIC (model 53230A, Keysight Corporation) is used to drive the reversal calculation unit, and further control the time delay unit (model DG645, Standard Research Systems Corporation). As a result, the one pulse per 10 milliseconds signal CLK-S can be time reversed and used to modulate the laser-S, whose wavelength is 1546.12 nm.

For the *Sync_Resp* process from server site to user site, the modulated optical signal propagates through the 230 km fiber link, and is filtered, amplified and detected at the user site. Enabled by the time reversal operation, the fiber propagation delay fluctuation in two directions can be eliminated. Measured by a TIC, the time difference between the detected time signal and the local CLK-U signal is used to drive the computing unit to compensate the clock offset of two sites. Furthermore, clock generators and time delay units at both sites are referenced to the synchronized 10 MHz atomic signal. This implementation is used to simulate fiber-based frequency dissemination, which has been developed rapidly in recent decades and has become quite mature and feasible [33]-[39]. In this way, CLK-U can be time synchronized to CLK-S. The synchronization result is characterized by TDEV and shown as the red line of Fig. 3. TDEV results of ~25 ps at 1 s and ~2 ps at 1000 s are obtained.

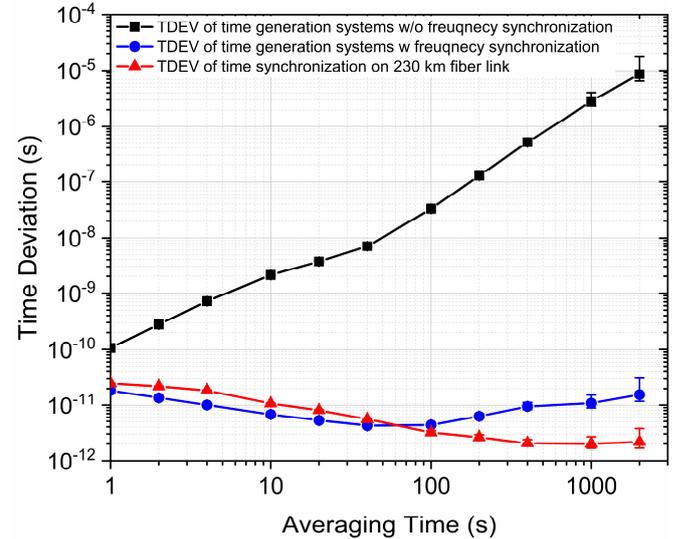

Fig. 3. TDEV results in different scenarios. The black line is the TDEV result of two time generation systems without frequency synchronization, the blue line is the TDEV result of two time generation systems with frequency synchronization, and the red line is the TDEV result of time synchronization on the 230 km fiber link.

To verify the importance of frequency synchronization for time synchronization, a direct way is to compare the TDEV of two time generation systems under the condition with/without frequency synchronization. Each time generation system consists of a clock generator and a time delay unit, without connecting any optical components or fiber. In the scenario without frequency synchronization, the TDEV result is plotted as the black line in Fig. 3. It is ~106 ps at an averaging time of 1 s and ~3 μs at an averaging time of 1000 s. In the scenario with frequency synchronization, two clock generators and two time delay units are all referenced to the external 10 MHz signals from the same frequency distribution amplifier. As shown by the blue line, the TDEV result is greatly improved, with ~19 ps at an averaging time of 1 s and ~11 ps at an averaging time of 1000 s obtained. This means that the frequency synchronization allows clock generators and time delay units at both sites to maintain a time consistency of one second, which can improve the time synchronization performance.

In addition, within the average time of ~60 s, the blue line is slightly lower than the red line. This means that the short-term performance of the time reversal enabled FOTS method is mainly limited by the specification of time generation systems. However, after the averaging time of ~60 s, the blue line starts to be higher than the red line. This is because even if frequency synchronization is achieved at both sites, there will be clock offset drift due to the fact that these two time generation systems cannot be identical. The clock offset drift can be compensated in real time through the time reversal enabled FOTS method. In our system, the compensation period is approximately 1 s. Since the blue line is higher than the red line after ~60 s, the compensation period is feasible within 60 s in our system. A longer compensation period can provide the possibility of synchronization for multiple remote user sites, with the help of time-division multiplexing technology and fast switchable optical switches. For example, if we carry out the time synchronization operation every 60 s for each user site, and the operation duration is within 1 s, theoretically, 60 user sites can be synchronized to 1 server site. Furthermore, if better time generation systems are used, the time synchronization performance of our system can be further improved. Moreover, the synchronization distance can be further extended, as the 230 km fiber connection has not significantly deteriorated the short-term TDEV of time generation systems according to the current results.

*B. Multiple-access time synchronization demonstration on a 230 km fiber link*

In the time reversal enabled FOTS method, it is the physical time signal that is transmitted in the fiber link. Hence, the evolution of the link delay can be continuously reflected on the time signal in real time, which makes it possible to realize multiple-access time synchronization. The principle of multiple-access time synchronization is shown in Fig. 4. There is an accessing node, which can be anywhere on the fiber link. The time signal transmitted from the user site is downloaded at the accessing node via an optical coupler as $t_{U-AN}$. At the server site, the operation of time reversal is the same as in Fig. 1(b). The time reversed signal is sent to the user site in the *Sync_Resp* process and then downloaded at the accessing node via an optical coupler as $t_{S-AN}$. Here, $t_{U-AN}$ can be formulated as

$$t_{U-AN} = t_{user} + \tau_{U-AN}, \quad (5)$$

where $\tau_{U-AN}$ is the propagation delay from the user site to the accessing node. $t_{S-AN}$ can be formulated as

$$t_{S-AN} = t_{s-reversal} + \tau_{S-AN}, \quad (6)$$

where $\tau_{S-AN}$ is the propagation delay from the server site to the accessing node and has the following relationship

$$\tau_{S-AN} + \tau_{U-AN} = \tau_{u-s}. \quad (7)$$

By substituting (2) and (7) into (6), we can have

$$t_{S-AN} = t_{server} + C + T_{offset} - \tau_{U-AN}, \quad (8)$$

The time difference $T_3$ between the two downloaded time signals at the accessing node is measured. According to (5) and (8), $T_3$ can be formulated as

$$T_3 = t_{S-AN} - t_{U-AN} = C + 2T_{offset} - 2\tau_{U-AN}. \quad (9)$$

Therefore, when delaying the time signal $t_{U-AN}$ by $T_3/2$, the synchronized time signal $t_{server} + C/2$ can be obtained at the accessing node.

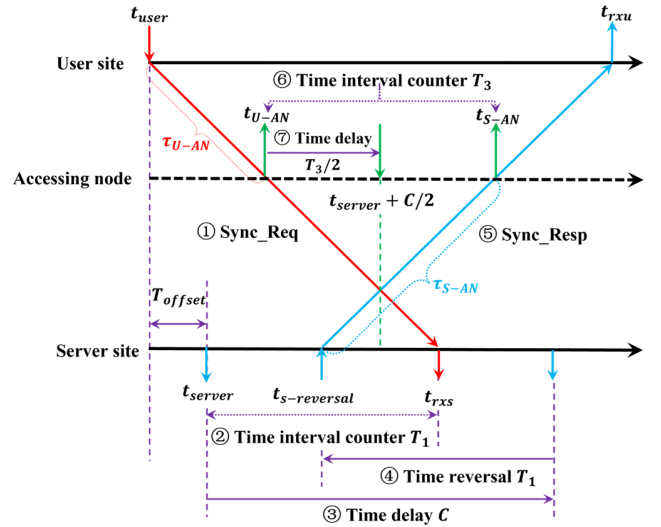

Fig. 4. Simplified timing model of multiple-access time synchronization based on time reversal. The upper and lower black axes indicate relative time relationships within one period.

The experimental setup of the multiple-access time synchronization test is shown in Fig. 5. The setup of the user site and server site is the same as the Fig. 2(b), and the fiber link is the same 230 km fiber link. For simplicity, we only show the clock signal CLK-U and the laser at the user site. The accessing node is 50 km away from the server site. A 2×2 fiber coupler is used to couple out the optical signals of two directions at the accessing node. After amplification and detection, the time difference of these two time signals is measured and sent to drive the computing unit for controlling the time delay unit to realize time synchronization.

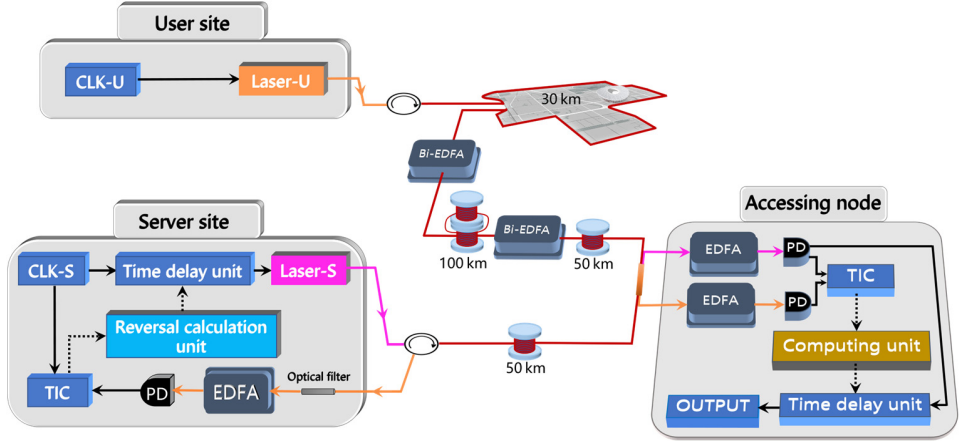

Fig. 5. The experimental setup of the multiple-access time synchronization on a 230 km fiber link. The accessing node is 50 km away from the server site.

Fig. 6 demonstrates the measured time difference and TDEV of the multiple-access time synchronization test. The time difference is plotted in Fig. 6(a), and the measurement duration is 10000 s. The TDEV is plotted in Fig. 6(b), with ~30 ps at 1 s and ~3 ps at 1000 s achieved. Although the accessing node in this multiple-access test is closer to the server site, its TDEV is slightly worse compared to the results in Fig. 3 (~25 ps at 1 s and ~2 ps at 1000 s). This is because the input signal of the time delay unit is the detected time signal from the fiber link, whose time jitter is worse than the time signal directly from the clock generator.

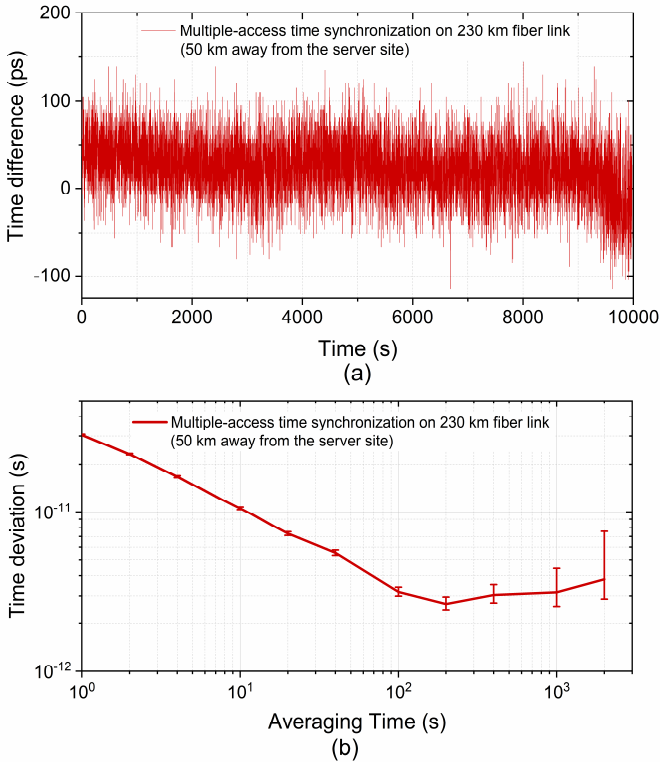

Fig. 6. The results of the multiple-access time synchronization test on a 230 km fiber link. (a) Time difference between the server site clock and the recovered time signal at the accessing node. (b) Time deviation of the multiple-access time synchronization test.

## IV. CONCLUSION

In this paper, we present an FOTS method enabled by time reversal to realize high-precision, low-complexity and scalable time synchronization, which will benefit the construction of the SAGIN in future B5G/6G scenarios and large-scale time synchronization network in next-generation radio telescope arrays. Compared with the commonly used two-way FOTS method, the proposed time reversal enabled FOTS method has three main advantages. First, the proposed method can calculate the clock difference of two locations without the need to measure or calculate the fiber link delay. Therefore, the date layer can be avoided, and the complexity of the system can be reduced significantly. Second, the synchronization process of this method is very simple, requiring only two steps: synchronization request and synchronization response. This process is a fairly direct and efficient way to achieve time synchronization. Third, it is feasible to achieve multiple-access time synchronization along the link, which is impossible for the traditional two-way FOTS.

The importance of frequency synchronization for time synchronization was also verified in this paper. With frequency synchronization, the TDEV result of our time generation system can be improved from ~106 ps @ 1 s and ~3 μs @ 1000 s to ~19 ps @ 1 s and ~11 ps @ 1000 s. Subsequently, the time synchronization test is conducted over a 230 km fiber link, which contains a 30 km urban link. TDEV results of ~25 ps at 1 s and ~2 ps at 1000 s are obtained. Since the physical time signal is transmitted in the fiber link, the proposed time reversal enabled FOTS method has the capability of multiple-access time synchronization. An accessing node 50 km away from the server site is chosen as a demonstration, with TDEV results of ~30 ps at 1 s and ~3 ps at 1000 s achieved.

This paper provides a demonstration of the proposed time reversal enabled FOTS method. If a better time generation system is used and the components at both server site and user site are designed to be integrated, the time synchronization performance will be further improved. In addition, the fiber link length of time synchronization can be further extended according to the current test results. The optimized design of the synchronization system and optical amplification is expected to advance the synchronization distance.

APPENDIX

Calibration is a necessary step in time synchronization. As shown in Fig. 7, the delay model of our system is characterized by fiber propagation delays and the hardware delays of two sites. The propagation delays of Bi-EDFA are considered last. At each site, there are transmission delays of the electrical-optical (E/O) converter and reception delays of the optical-electrical (O/E) converter. $\tau_{TXS}$ and $\tau_{RXS}$ are the transmission delay and reception delay for the server site, respectively, while $\tau_{TXU}$ and $\tau_{RXU}$ are the transmission delay and reception delay for the user site, respectively, These hardware delays can be regarded as constant, involving the delays of electronic components, DFB lasers, EDFAs, PDs and internal cables, etc. $\tau_{u-s}$ and $\tau_{s-u}$ are the one-way propagation delays of the fiber link in the *Sync_Req* direction and *Sync_Resp* direction, respectively. Hence, the measurement result $T_1$ of the TIC-S at the server site can be specifically described as

$$T_1 = \tau_{TXU} + \tau_{u-s} + \tau_{RXS} - T_{offset}, \quad (10)$$

where $T_{offset}$ is still the clock offset of two sites. The measurement result $T_1$ is input to a reversal calculation unit. Subsequently, the value $C - T_1$ is calculated and input to the time delay unit to generate a reversed time signal. $C$ is a known constant value. Due to the nonideality of the time delay unit, there exists a deviation $\tau_{delay\_s}$ between the actual delay value and the input value of the time delay unit at the server site. Consequently, the measurement result $T_2$ of TIC-U at the user site can be expressed as

$$T_2 = T_{offset} + C - T_1 + \tau_{TXS} + \tau_{s-u} + \tau_{RXU} + \tau_{delay\_s}. \quad (11)$$

By substituting (10) into (11), we can have

$$T_2 = 2T_{offset} + C + \tau_{TXS} + \tau_{RXU} - \tau_{TXU} - \tau_{RXS} + \tau_{delay_s} + \tau_{s-u} - \tau_{u-s}. \quad (12)$$

According to (12), to obtain the clock offset $T_{offset}$, it is necessary to calibrate the hardware delay $\tau_{HD}$ and fiber propagation delay asymmetry $\tau_{FPDA}$, where $\tau_{HD}$ can be expressed as

$$\tau_{HD} = \tau_{TXS} + \tau_{RXU} - \tau_{TXU} - \tau_{RXS} + \tau_{delay\_s}, \quad (13)$$

and $\tau_{FPDA}$ can be expressed as

$$\tau_{FPDA} = \tau_{s-u} - \tau_{u-s}. \quad (14)$$

To calibrate the hardware delay, the server site and user site are directly connected with an optical attenuator. In this way, the O/E converter can avoid receiving excessive optical power and the fiber propagation delay can be neglected. Then, the time reversal function of the system is enabled to realize time synchronization between two sites. At this point, assuming that the initial value measured by TIC-U is $T_{2\_INIT}$ and the initial clock offset is measured as $T_{offset\_INIT}$ in advance, the hardware delay $\tau_{HD}$ can be calibrated as

$$\tau_{HD} = T_{2\_INIT} + 2T_{offset\_INIT} - C. \quad (15)$$

The delay deviation $\tau_{delay\_u}$ of the time delay unit at the user site also needs to be calibrated. This delay deviation is easy to calibrate through using a TIC to measure the time difference between the input time signal and the output time signal.

In long-distance transmission over a fiber link, to suppress Rayleigh backscattering, the laser wavelengths at two sites are normally different. As a result, there is a fiber propagation delay asymmetry $\tau_{FPDA}$ in two directions caused by chromatic dispersion. This delay difference can be calculated as

$$\tau_{FPDA} = \tau_{s-u} - \tau_{u-s} = (\lambda_2 - \lambda_1)D_A, \quad (16)$$

where $\lambda_1$ is the laser wavelength at the server site, $\lambda_2$ is the laser wavelength at the user site and $D_A$ is the accumulated dispersion of the fiber link. If the entire fiber link with a length of $L$ has the same dispersion coefficient $D$, the accumulated dispersion can be calculated as $D_A = DL$. Another method is to use a chromatic dispersion analyzer to measure the accumulated dispersion [35]. The laser wavelength difference can be measured by an optical spectrum analyzer.

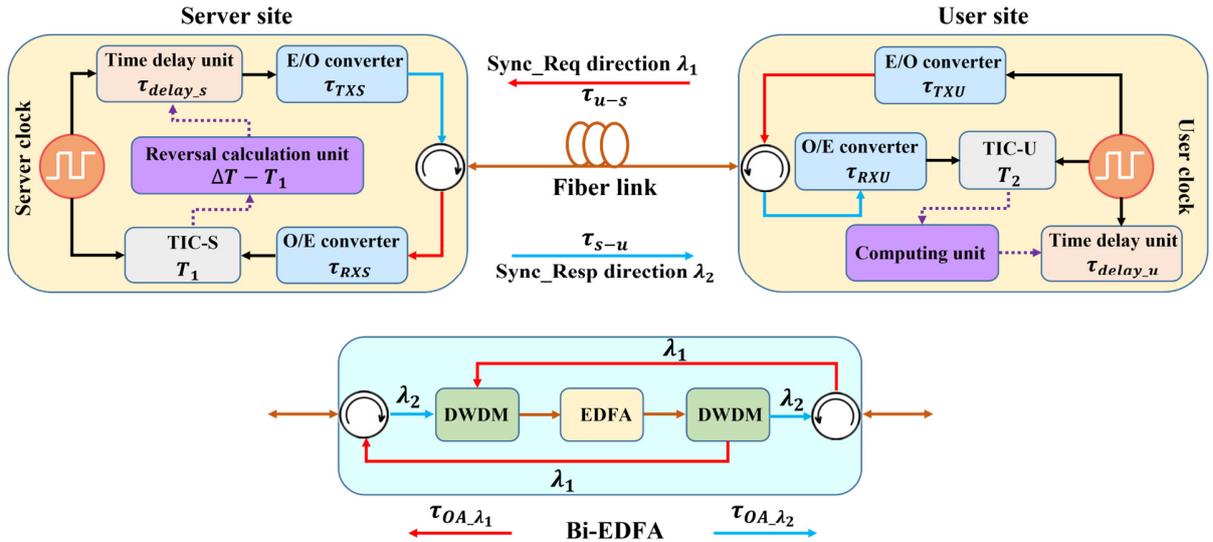

Fig. 7. The delay model of the time reversal enabled FOTS system, including the hardware delays at both server site and user site, fiber propagation delays and propagation delays of Bi-EDFA. E/O converter, electrical-optical converter; O/E converter, optical- electrical converter; DWDM, dense wavelength division multiplexer.

Another important step is the calibration of the Bi-EDFA in long-distance transmission. In [35], a single-path bidirectional amplifier is used to achieve good inherent symmetry, but the gain is limited due to backscattering. The two-path scheme with a single EDFA on each path can effectively solve the problem of gain limitation [39], but it is difficult to obtain good symmetry. To realize a trade-off between the gain and symmetry, two optical circulators and two dense wavelength division multiplexers (DWDMs) are used to form bidirectional propagation, as shown in Fig. 7. The DWDM can suppress Rayleigh backscattering. Only one EDFA is used to achieve optical amplification in two directions. In this way, the path of optical amplification in both directions can be ensured to be consistent. It is not difficult to calibrate the propagation asymmetry of the DWDM-based Bi-EDFA, which can be expressed as $\tau_{OAD} = \tau_{OA\_\lambda_1} - \tau_{OA\_\lambda_2}$.

In this way, we have obtained the hardware delay $\tau_{HD}$, the fiber propagation delay asymmetry $\tau_{FPDA}$ and the propagation asymmetry of Bi-EDFA $\tau_{OAA}$. Hence, $T_2$ can be re-expressed as

$$T_2 = C + \tau_{HD} + \tau_{FPDA} + \tau_{OAA} + 2T_{offset}. \tag{17}$$

The actual clock offset $T_{offset}$ of two sites can be calculated as

$$T_{offset} = 0.5(T_2 - C - \tau_{HD} - \tau_{FPDA} - \tau_{OAA}). \tag{18}$$

REFERENCES

[1] J. Ballato, "Optical fifiber: Through the looking glass," *Opt. Photon. News*, vol. 33, no. 1, pp. 32–40, 2022.

[2] Z. Jiang *et al*., "Comparing a GPS time link calibration with an optical fibre self-calibration with 200 ps accuracy," *Metrologia*, vol. 52, no. 2, pp. 384–391, 2015.

[3] P. Loschmidt *et al*., "White rabbit - sensor/actuator protocol for the CERN LHC particle accelerator," in *Proc. IEEE*, Christchurch, New Zealand, 2009.

[4] M. Antonello *et al*., "Precision measurement of the neutrino velocity with the ICARUS detector in the CNGS beam," *J. High Energy Phys.*, vol. 2012, no. 11, p. 49, 2012.

[5] C. Clivati *et al*., "Common-clock very long baseline interferometry using a coherent optical fiber link," *Optica*, vol. 7, no. 8, pp. 1031-1037, 2020.

[6] P. Krehlik *et al*., "Fibre-optic delivery of time and frequency to VLBI station," *Astronomy & Astrophysics*, vol. 603, p. A48, 2017.

[7] B. Shillue *et al*., "The ALMA photonic local oscillator system," in Proc. SPIE, Amsterdam, Netherlands, 2012.

[8] S. T. Garrington *et al*., "e-MERLIN," in *Proc. SPIE*, Glasgow, Scotland, 2004.

[9] S. W. Schediwy *et al*., "The mid-frequency Square Kilometre Array phase synchronisation system," *Publications Astronomical Soc. Aust.*, vol. 36, p. e007, 2019.

[10] M. Jiménez-López *et al*., "A Fully Programmable White-Rabbit Node for the SKA Telescope PPS Distribution System," *IEEE Trans. Instrum Meas.*, vol. 68, no. 2, pp. 632-64, 2022.

[11] R. Selina *et al*., "The Next-Generation Very Large Array: a technical overview," in *Proc. SPIE*, Austin, TX, USA, 2018.

[12] M. Calhoun, S. Huang, and R. L. Tjoelker, "Stable photonic links for frequency and time transfer in the deep-space network and antenna arrays," in *Proc. IEEE*, vol. 95, no. 10, pp. 1931–1946, Oct. 2007.

[13] Y. Liang *et al*., "IEEE 1588-Based Timing and Triggering Prototype for Distributed Power Supplies in HIAF," *IEEE Trans. Instrum Meas.*, vol. 71, pp. 1-9, 2022, Art. no. 5502309.

[14] A. Niell *et al*., "Demonstration of a Broadband Very Long Baseline Interferometer System: A New Instrument for High-Precision Space Geodesy," *Radio Sci.*, vol. 53, no. 10, pp. 1269–1291, 2018.

[15] C. Lisdat *et al*., "A clock network for geodesy and fundamental science," *Nature Commun.*, vol. 7, no. 1, p. 12443, 2016.

[16] L. Sliwczynski *et al*., "Fiber-Based UTC Dissemination Supporting 5G Telecommunications Networks," *IEEE Commun. Mag.*, vol. 58, no. 4, pp. 67–73, 2020.

[17] Z. Zhang *et al*., "Time Synchronization Attack in Smart Grid: Impact and Analysis," *IEEE Trans. Smart Grid*, vol. 4, no. 1, pp. 87–98, 2013.

[18] K.-K.-R. Choo, Z. Yan, and W. Meng, ''Editorial: Blockchain in industrial IoT applications: Security and privacy advances, challenges, and opportunities,'' *IEEE Trans. Ind. Informat.*, vol. 16, no. 6, pp. 4119–4121, 2020.

[19] K. Fan, S. Wang, Y. Ren, K. Yang, Z. Yan, H. Li, and Y. Yang, "Blockchain-Based Secure Time Protection Scheme in IoT," *IEEE Internet Things J.*, vol. 6, no. 3, pp. 4671–4679, 2019.

[20] L. Narula and T. E. Humphreys, "Requirements for Secure Clock Synchronization," *IEEE J. Sel. Topics Signal Process.*, vol. 12, no. 4, pp. 749–762, 2018.

[21] S. Wan *et al*., "Fair-Hierarchical Scheduling for Diversified Services in Space, Air and Ground for 6G-Dense Internet of Things," *IEEE Trans. Netw. Sci. Eng.*, vol. 8, no. 4, pp. 2837–2848, 2021.

[22] S. Dang *et al*., "What should 6G be?" *Nature Electron.*, vol. 3, pp. 20–29, 2020.

[23] J. Kodet, P. Pánek, and I. Procházka, "Two-way time transfer via optical fiber providing subpicosecond precision and high temperature stability." *Metrologia*, vol. 53, no. 1, pp. 18–26, 2016.

[24] S.-C. Ebenhag *et al*., "Measurements and Error Sources in Time Transfer Using Asynchronous Fiber Network." *IEEE Trans. Instrum Meas.*, vol. 59, no. 7, pp. 1918–1924, 2010.

[25] F. Hou *et al*., "Fiber-optic two-way quantum time transfer with frequency-entangled pulses", *Phys. Rev. A*, vol. 100, no. 2, 2019.

[26] M. Rost *et al*., "Time transfer through optical fibres over a distance of 73 km with an uncertainty below 100 ps." *Metrologia*, vol. 49, no. 6, pp. 772–778, 2012.

[27] O. Lopez *et al*., "Simultaneous remote transfer of accurate timing and optical frequency over a public fiber network," *Appl. Phys. B*, vol. 110, no. 1, pp. 3–6, 2013.

[28] F. Zuo *et al*., "Multiple-Node Time Synchronization Over Hybrid Star and Bus Fiber Network Without Requiring Link Calibration", *J. Lightw. Technol.*, vol. 39, no. 7, pp. 2015–2022, 2021.

[29] Y. Xu and L. V. Wang, "Time Reversal and Its Application to Tomography with Diffracting Sources," *Phys. Rev. Lett.*, vol. 92, no. 3, 2004.

[30] G. F. Edelmann *et al*., "An initial demonstration of underwater acoustic communication using time reversal," *IEEE J. Ocean Eng.*, vol. 27, no. 3, pp. 602–609, 2002.

[31] E. Kerbrat *et al*., "Ultrasonic nondestructive testing of scattering media using the decomposition of the time-reversal operator," *IEEE Trans. Ultrason. Ferr. Freq. Contr.*, vol. 49, no. 8, pp. 1103–1113, 2002.

[32] M. Fink, "Time reversal of ultrasonic fields. I. Basic principles", *IEEE Trans. Ultrason. Ferr. Freq. Contr.*, vol. 39, no. 5, pp. 555–566, 1992.

[33] B. Wang et al., "Precise and continuous time and frequency synchronisation at the $5 \times 10^{-19}$ accuracy level," *Sci. Rep.*, vol. 2, p. 556, 2012.

[34] Y. He et al., "Long-distance telecom-fifiber transfer of a radio-frequency reference for radio astronomy," *Optica*, vol. 5, no. 2, pp. 138–146, 2018.

[35] L. Sliwczynski *et al*., "Dissemination of time and RF frequency via a stabilized fibre optic link over a distance of 420 km," *Metrologia*, vol. 50, no. 2, pp. 133–145, 2013.

[36] D. X. Wang *et al*., "Stable radio frequency dissemination via a 1007 km fiber link based on a high-performance phase lock loop," *Opt. Exp.*, vol. 26, no. 19, pp. 24479–24486, 2018.

[37] K. Predehl et al., "A 920-kilometer optical fiber link for frequency metrology at the 19th decimal place," *Sci.*, vol. 336, no. 6080, pp. 441–444, 2012.

[38] C. Gao et al., "Fiber-based multiple-access ultrastable frequency dissemination," *Opt. Lett.*, vol. 37, no. 22, pp. 4690–4692, 2012.

[39] Y. Chen et al., "Long-Haul High Precision Frequency Dissemination Based on Dispersion Correction," *IEEE Trans. Instrum Meas.*, vol. 71, pp. 1-7, 2022, Art. no. 5503207.

**Yufeng Chen** received the B.S. degree in 2018 from Tsinghua University, Beijing, China, where he is currently working toward the Ph.D. degree in instruments science and technology. His current research focuses on fiber-based time-frequency synchronization and its applications.

**Hongfei Dai** received the B.S. degree in 2020 from Tsinghua University, Beijing, China, where he is currently working toward the Ph.D. degree in instruments science and technology. His current research focuses on fiber-based time-frequency synchronization.

**Wenlin Li** received the B.S. degree in 2022 from China University of Geosciences, Beijing, China. He is currently working toward the Ph.D. degree in instruments science and technology in Tsinghua University, Beijing, China. His current research focuses on phase noise analysis and fiber-based time-frequency synchronization.

**Fangmin Wang** received the B.S. degree in 2017 from Shanxi University, Taiyuan, China. He is currently working toward the Ph.D. degree in instruments science and technology in Tsinghua University, Beijing, China. His current research focuses on distributed and real-time timekeeping network.

**Bo Wang** (Senior Member, IEEE) received the Ph.D. degree in optics from Shanxi University, Taiyuan, China, in 2007. From 2007 to 2010, he was a Postdoctoral Researcher with the Max-Planck Institute for the Science of Light, Erlangen, Germany. He is currently a Tenured Associate Professor with the Department of Precision Instrument, Tsinghua University, Beijing, China. His current research interests include space-time and standards technology, and fiber network sensing. He is also a senior member of OPTICA.

**Lijun Wang** received the Ph.D. degree in physics from the University of Rochester, Rochester, NY, USA, in 1992. From 2004 to 2010, he was a Professor with the Department of Physics, University of Erlangen-Nuremberg, Erlangen, Germany, and the Director of Max-Planck Research Group, Institute for Optics, Information and Photonics, Erlangen, Germany. He is currently a Professor with Tsinghua University, Beijing, China. His current research interests include space-time and standards technology. He is also a Fellow of OPTICA.